**Deriving general relativity from considerations on quantum information**

Thomas Görnitz

(FB Physik, Goethe-Universität, Frankfurt/Main)

goernitz@em.uni-frankfurt.de

**Introduction**

For a long time a lot of work has been done to quantize gravitation. The starting point for such considerations was almost ever general relativity. This is a structure of very high mathematical beauty and complexity. It was used as basic structure that should be quantized in one or another way, like other field theories. But up to now, all such efforts have remained without convincing success.

During this process it has become obvious that "more and more of the same" does not help to show the way. The door from relativity theory to quantum theory could not be opened until now. However, by making a step backward, it becomes obvious that it is possible to open the door in the other direction, from quantum theory to gravitation.

Such a step backwards allows a wider view. It could be initiated by a new thinking on the concept of "matter" as well as by a reflection about general relativity. Some arguments speak for such a change in the procedure which has been used for a long time. They have an epistemological aspect and are related to assumptions that do not seem to have been considered so far. Therefore, they have to reach beyond pure mathematical considerations.

A proposal to derive Einstein's equations from quantum information was presented by Görnitz[1]. That there is a close connection between quantum information and gravity was shown by him a long time ago[2] on the basis of the Bekenstein-Hawking-Entropy[3], but such ideas only recently gained attention. Today such an idea seems to be no longer an outsider concept. "… the central notion needed to derive gravity is information"[4] – such a sentence was far away mainstream some years ago.

The basic step for our consideration is the introduction of an abstract quantum information, that means an information which is primarily free of special meaning. For daily life information is almost ever identified with meaning and due to the fact that the abstract quantum information has no specific meaning a priory, a new denotation was introduced: Protyposis.

---

[1] Görnitz, Th. 2009
[2] Görnitz, Th. 1988a,b; Görnitz, Ruhnau 1989
[3] Bekenstein 1973,1974, Hawking 1973
[4] Verlinde 2010

By few assumptions, for which good physical reasons can be given, consequences of the theory are on the one hand a realistic cosmology with a homogeneous and isotropic background, on the other hand, from these propositions a state equation results for the energy density and pressure of this quantum information.

Today this proposed cosmological model is in a better agreement with the observational data than at the time of its first presentation in 1988.[5] This model solves the so-called cosmological problems and it can explain the so-called "dark energy" as well.

This cosmology is established in an abstract way and based on quantum theory of the Protyposis. If the cosmologically established proportionality between energy-momentum-tensor and Einstein tensor is required to remain valid for local variations of density and pressure as well, then the Einstein equations follow.

This new way breaks with some of the epistemological postulates which in the context of general relativity were seen as unalterable. Nevertheless, the fundamental importance of this theory remains unaffected for all the mathematical approximations developed from general relativity for a description of localized gravitational effects.

**Relations between general relativity, quantum theory and cosmology**

Einstein had written about general relativity theory (GR) that it looks like a building with one wing of marble (the geometrical part of the equation, the left one) and the other wing of wood (the material part of the equation, the right one).[6] However, the sense of an equation is that both parts are equal. If the right wing is of wood, then according to the idea of an equation, the left one likewise. Einstein had created a picture that is not really appropriate.

Of course, the geometrical relations in Einstein's equations are of an unsurmountable clarity. However, contrary to any pure mathematical problem, in physics the connection to experience is absolutely essential. Since the discovery of quantum theory it has become obvious that the wonderful mathematical models of space-time manifolds and geodesic lines become invalid if the considerations reach such a degree of accuracy that quantum theory must be included. Because the relations between geometrical models and reality are not so unambiguous, it seems to be useful to scrutinize all the efforts which has been made so far to establish the transition between gravitational theory and quantum phenomena. It is time to reflect whether a quantization of GR in one or the other form is the right way to connect these fields of physics.

---

[5] Görnitz 1988 a, b
[6] Einstein, 1984, p. 90

The quantum structure was imposed on physics only after physics had achieved a very high degree of precision in its description of nature. Only when the precision becomes very high, quantum effects happen to be essential. Therefore, it is obvious that not for every mathematical structure a quantized form has to exist, even if it models natural phenomena in an adequate way. Only for such mathematical models a quantized form will exist which are especially well-fit to nature and which can be seen as fundamental.

From this point of view we will have a look at GR.

Generally speaking, the implication of an equation is equivalent to the set of its solutions and vice versa. Given the equation one has, in principle, all of its solutions. And, again in principle, having all of its solutions, the equation can be reconstructed.

Between GR and all other equations in physics there is a fundamental difference. If a solution in GR is not used only as an approximation, then in every case, it describes a cosmos in its full spacelike und timelike extensions.

As we have said already, in contradiction to mathematics, physics needs its connection to experience. One can define the universe as the set of all possible phenomena for which it is not impossible by first principles to get at one time or another some empirical knowledge. The cosmos is therefore the totality of all possible empirical accessible physical phenomena. But then, by this definition, the cosmos, as an item of physics, is with necessity a single one. A plural for cosmos is impossible in the range of physics. Of course, it is possible to speak about a multitude of universes in mathematics, in science fiction or in fantasy, but not in empirical based physics. The only certain knowledge we can have on such hypothetical universes is, that every empirical knowledge is impossible.

Einstein's equations possess infinitely many solutions. At best, only one of these can describe the real cosmos of our physical empirical experience. In a strong sense all the other solutions can not be related to physical reality if not taken only as approximations. Therefore, almost all solutions of these equations are pure mathematics. So, in these equations too much is stated that belongs only to mathematics and no more to physics. Having this in mind it can be considered as a further hint reflecting the quantization of Einstein's equations.

On the other hand, the experimental and observational checks of GR, respectively its linear approximations, are overwhelmingly good. Therefore no attempt which wants to declare these equations **not** as a very good description of observations and experiences can be justified. Nevertheless, they have to be interpreted as an approximation, of course.

The aim of connecting cosmology and physics has further aspects which are seldom verbalized in physical literature, for example the problem of empirical foundation.

**The problem of empirical foundation**

In current cosmology we find some so-called "cosmological problems". For instance, the horizon-problem is connected with the high homogeneity of background radiation, which even comes from distant areas which, by the models taken into account, could not have had any causal connection. The problem of the "cosmological constant" is connected with the observation that in many models an extremely little term is needed, which nevertheless is not zero. The "flatness-problem" states that space is nearly flat. Therefore, many cosmological models make, for simplicity, the space Euclidian, but then the problem of empirical foundation arises.

Of course, nothing can be said against the possibility to understand cosmology as a part of pure mathematics and therefore to examine the cosmological solutions of GR with all of its interesting mathematical properties. It is not necessary that all of these have something to do with physical reality. On the other hand, many physicists and astronomers use no cosmological models and restrict themselves on the inspection of the metagalaxis. They do not have the demand for an examination of the cosmos as a whole. But only if the metagalaxis represents a fundamental part of the cosmos, will it have meaning for the whole, and the assertions about the whole only have a weak foundation if the unseen part is essentially larger then the observed one.

There are good reasons to see cosmology as a part of physics, but then the problem of empirical foundation arises. If cosmological statements are intended to be part of physics, then it is not possible to restrict them to the metagalaxis and otherwise to postulate a flat Euclidean space. With a flat and actual infinite position space all related cosmological models became physically worthless. For this case all the empirical data trivially cover exactly zero percent of the whole, and this is insufficient, at least in physics.

In the last decade wonderful cosmological data was found. If cosmology should be meaningful connected with these data, then an actual finite volume of the cosmic space is a necessary precondition. An inference from zero percent to hundred percent is difficult in any science, whereas if the data cover perhaps more than half of the whole, as it would be for a finite cosmic volume, then there are good reasons to believe in the extrapolating conclusions.

The condition of a finite space cannot be compensated by a "cosmological principle" which states that an infinite cosmos is always like this accidental part that we see, because this principle would be likewise founded on zero percent of the empirical data.

**Quantum theory is universal**

In connection with cosmological questions a popular misunderstanding becomes relevant which states quantum theory as a "theory of microphysics".

At present the physicists agree that quantum phenomena become unavoidable in approaching the temporal neighborhood of the big bang with its extreme high densities and temperatures in its small volume. This agreement is connected with the impression that quantum theory is "a theory of the small". Of course, it is correct that without quantum theory nothing can be understood in the area of the small things. The reason is that quantum theory as the "physics of precision" is unavoidable in the range of the small objects. For large objects the precision of quantum theory is not always needed. Therefore these can be treated with the means of the not so precise classical physics.

To avoid misunderstandings it is useful to remember that classical physics predicts a deterministic structure for the evolution of facts by ignoring the relational structure of reality. But such a prediction is not true. Quantum theory, as the "physics of relations", predicts in a more accurate way a deterministic structure only for the evolution of the possibilities, whereas the coming facts are not strictly determined.

The better accuracy of quantum theory is necessary sometimes, even for large systems. So quantum theory seems unavoidable for an understanding of the groundstate of a system, even if quantum theory cannot be used always like a recipe. An example is the cosmological term $\Lambda$ in Einstein's equations. It is understood mostly as a constant. Schommers gives an overview worth reading on the problems connected with $\Lambda$.[7]

In the later development of the theory $\Lambda$ was connected with the "energy of the vacuum", because such a term represents the properties of the energy-momentum-tensor of the vacuum in Minkowski-space. But a computation of $\Lambda$ by methods of quantum field theory became wrong in many orders of magnitude. This result is not so surprising since the introduction of such a constant is arguable with good quantum physical reasons. All the empirical data refer to an expansion of cosmic space, and one of the first experiences in quantum theory is that the ground state energy of a system, its vacuum energy, depends on the extension of its spacelike volume. Therefore it cannot be a constant if the space expands.

---

[7] Schommers, 2008

It will be shown in which way the problem can be solved by considerations from abstract quantum information.

**A cosmological model from abstract quantum theory**

The considerations above have shown that a connection between quantum theory and general relativity is not forced to start with Einstein's equations and then trying to quantize these. As an alternative it is possible to derive a cosmological model from a theory of abstract quantum information. By the plausible assumption that the first law of thermodynamics is valid, results a global relation between energy density and pressure. The consequential energy-momentum-tensor has at a first view no relation to the metric of the model, but taking the derivations of the metric into the considerations, then the resulting Einstein-tensor is proportional to this energy-density-tensor.

It seems natural to propose the condition that this relation between the two tensors should remain valid also for local deviations from the homogeneous distribution of energy density and pressure. Then Einstein's equations follow as an approximation for local variances of the cosmological situation.

In a series of papers such a cosmological model, constructed from quantum theoretical considerations, was presented.[8] It describes a closed cosmic space, expanding with velocity of light.

At the time of its presentation the cosmological folklore was convinced that a closed cosmos is recollapsing, and it is not possible for it to expand forever. Later on, the notion became popular that the universe is flat and not closed. Therefore this model was not very popular, even if it appeared in a textbook some years later.[9]

In the meantime, quantum information is no longer considered just by some theoreticians. On the contrary, the importance of quantum information and its properties becomes visible in a rising number of successful exciting experiments.[10] And in the last time even a connection from information to gravitation has become conceivable.[11]

---

[8] Görnitz, 1988 a, b; Görnitz, Ruhnau, 1989
[9] Goenner, 1994, S. 87

The suggested cosmological model is not only unconcerned about the problems related to the horizon, the flatness or the cosmological constant; moreover, it becomes obvious that it fits better to the empirical data of today[12] then to the data that were acknowledged at the time of its first presentation.

The considerations given here are underpinned by an important paper of Jacobson.[13] He showed that Einstein's equations can be derived from the thermodynamics of black holes:

"The four laws of black hole mechanics, which are analogous to those of thermodynamics, were originally derived from the classical Einstein equation. With the discovery of the quantum Hawking radiation, it became clear that the analogy is, in fact, an identity. How did classical general relativity know that the horizon area would turn out to be a form of entropy, and that surface gravity is a temperature? In this Letter I will answer that question by turning the logic around and deriving the Einstein equation from the proportionality of entropy and the horizon area together with the fundamental relation $\delta Q = T\, dS$ connecting heat $Q$, entropy $S$, and temperature $T$. Viewed in this way, the Einstein equation is an equation of state. It is born in the thermodynamic limit as a relation between thermodynamic variables, and its validity is seen to depend on the existence of local equilibrium conditions. This perspective suggests that it may be no more appropriate to quantize the Einstein equation than it would be to quantize the wave equation for sound in air."

Jacobson's derivation rests on thermodynamics and Hawking-radiation. In the following it will be shown that the Hawking-radiation is an expression for the entropy of space-time, or more clearly, for the quantum information that is the fundament for space-time altogether. So it becomes possible to start from the fundamental abstract quantum information instead of its special case of Hawking radiation.

In the 1950ies C. F. v. Weizsäcker proposed that physics should be founded on an abstract quantum information - Ur-alternatives.[14] Later on, the idea of fundamentality of quantum information was also seen for instance by Finkelstein[15] and Wheeler.[16]

---

[10] see e.g. papers of Zeilinger, Weinfurtner and many others.
[11] Padmanabhan, 2010, Verlinde 2010
[12] Riess et. al, 2004, Görnitz, 2006, Görnitz & Görnitz, 2008, p. 148
[13] Jacobson, 1995
[14] Weizsäcker, 1955, 1958, 1971, 1985; Scheibe et al., 1958
[15] Finkelstein 1968
[16] Wheeler 1990

That abstract quantum information is a reasonable assumption for a basic of physics became obvious at least since Görnitz, Graudenz and v. Weizsäcker showed in which way a relativistic quantum particle, this means an irreducible representation of the Poincaré group, can be constructed from qubits.[17] Such particles can be described in a framework of second quantization by Para-Bose creation and annihilation operators of qubits.

Weizsäcker had postulated that the three dimensions of position space are the consequence of the SU(2)-symmetry of the qubits. Later on, Drieschner[18] elaborated this idea further on. Unfortunately, their model contradicted general relativity, which hindered its acceptance.

**Abstract quantum information - Protyposis**

Weizsäcker had postulated "an »absolute« conception of information is meaningless"[19]. Therefore, his Ur-alternatives could not be understood as absolute entities. It became necessary to go out of this dictum from Weizsäcker and to understand the conception of "information" in such an abstract way that neither emitter and receiver nor any concrete meaning are kept in mind. Only such a far reaching abstraction allows quantum information to become absolute and then also equivalent to matter and energy.

With arguments from group theory and thermodynamics of black holes, Görnitz[20] was able to derive a cosmological model from considerations of quantum information, that is not in contradiction to general relativity. The abstract quantum information gets a connection to cosmology and black hole thermodynamics. In this process the quantum information lost every special meaning, therefore it was named with the new term "Protyposis". The new name is used to crack the immediate association of information with meaning.

This new name is important also to illuminate the difference between information and entropy. Only entropy can be measured in physics but only a part of the information is inaccessible and therefore to classify as entropy. To see how much qubits a particle "is", all the information, the accessible and the inaccessible, is to taken into account. Therefore Görnitz proposed a gedankenexperiment[21] that transduces all of the information into entropy. Such a gedankenexperiment consists of a huge black hole which encloses all the matter of the universe, and which has the same extension as the corresponding Friedman-Robertson-Walker-universe. If in this gedankenexperiment the particle is put into the black hole then no information of it is accessible any more and so all of its information is converted to entropy and becomes therefore measurable.

---

[17] Görnitz, Graudenz, v. Weizsäcker, 1992; Görnitz, Schomäcker, 1996
[18] Drieschner, 1979
[19] v. Weizsäcker, 1985, S. 172
[20] Görnitz, 1988
[21] Görnitz, 1988

To localize a small particle in the huge cosmic space requires a huge amount of information which is not totally hidden. For example, the tiny Planck-black-hole has a horizon of one Planck-length squared and so by definition an entropy of one bit, but if it is put into the black hole of the gedankenexperiment then the entropy of the black hole increases by approximately $10^{60}$ bits.

**Postulates for cosmology**

If the container of the reality, the real space-time of all our experiences and observations, is not a special solution of a general law (what is the meaning of all of its other unreal solutions?) then we have to go an alternative way to cosmology.

With some simple arguments a cosmological model can be derived from the abstract quantum information, from the Protyposis. The postulates are:

1. Abstract quantum information constitutes a basis for physics.
2. The energy of a quantum system is proportional to the inverse of its wave length.
3. For a closed system the first law of thermodynamics applies: *dU + pdV=0*
4. There is a distinguished velocity: *c*

**The introduction of position space**

If abstract quantum information constitutes a basis for physics then all physical phenomena must be representable by qubits. Then physical reality can be represented in the tensor product of a huge number of representations of a qubit.

The state space of a qubit is the two-dimensional complex space $\mathcal{C}^2$. The symmetry group, which leaves invariant the absolute value of the scalar product in this space, contains the group SU(2), and the phase transformations of the U(1) and complex conjugation.

The essential group is SU(2); its maximal homogeneous space is a $\mathcal{S}^3$. It is the three-dimensional surface of a four-dimensional sphere. We specify Weizsäcker's idea and postulate: The position space of physics and astronomy is represented by this $\mathcal{S}^3$.

The parameter of the U(1) is a formal time. But if a rising number of qubits is put under considerations then there is no more a unitary evolution and the U(1) has to be changed into a $\mathbb{R}^+$.

The Hilbert space of the quadratic integrable functions on this $\mathbb{S}^3$ is the carrier space for the regular representation of the SU(2). Every irreducible representation of the SU(2) is contained in the regular representation.

The states of a single qubit are represented by a two-dimensional representation of the SU(2). This representation as a subrepresentation of the regular representation is spanned by such functions that divide the $\mathbb{S}^3$ in only two halves. It is analogous to the sinus function in one dimension which divides the circle in two halves. A quantum theoretical combination of many qubits creates the tensor product of the two-dimensional representations. This can be reduced into irreducible representations of higher dimensions and in these representations there are functions with much better localization. (see Fig. 1)

This concept is in some sense complementary to the so-called holographic principle. Since the times of the old Greeks there has been a tendency in science to divide the physical reality into smaller and smaller parts. In the present these are atoms, elementary particles und the intended strings. This tendency is continued with the holographic principle: "The most important assumption will be that the information associated with a part of space obeys the holographic principle."[22] In it a bit is represented by an area of Planck length squared, therefore by the smallest existing area in the universe.

Inverse to the holographic principle the Protyposis concept is based on the insight that a quantum bit is of a cosmic dimension, so its "wave function" is extended over the whole cosmic space, whereas a very strong localization requires a huge amount of quantum information. So, this concept does not any more search for the simple in the little ones. For the Protyposis a natural and undissolvable connection between cosmology and quantum theory is unavoidable. It makes obvious that space and time in its wholeness and the material contend in it are consequences of the abstract quantum information. The holographic principle works on a two-dimensional surface in space, Protyposis takes the whole space into the considerations.

That gravity is a consequence of information can be seen in both approaches, in Verlindes and in the Protyposis conception. In a recent paper Hossenfelder[23] makes clear that the path from

---

[22] Verlinde 2010
[23] Hossenfelder 2010

Newtonian gravity to information can be taken in both directions and also that the holographic principle is not as essential as it seems to be.

I hope that some of Hossenfelder's questions about the relevance of quantum information for general relativity are answered in the present paper.

The Protyposis concept with its natural association to the cosmos has the benefit that in it a natural system of measurement can be given, whereas on the holographic way the value of ℏ remains indeterminate.

A one-dimensional example for the construction of something localized out of many extended things can be provided with the sinus function. Whereas the sinus divides an interval into two extended areas, the products of many of these functions can produce extremely localized states.

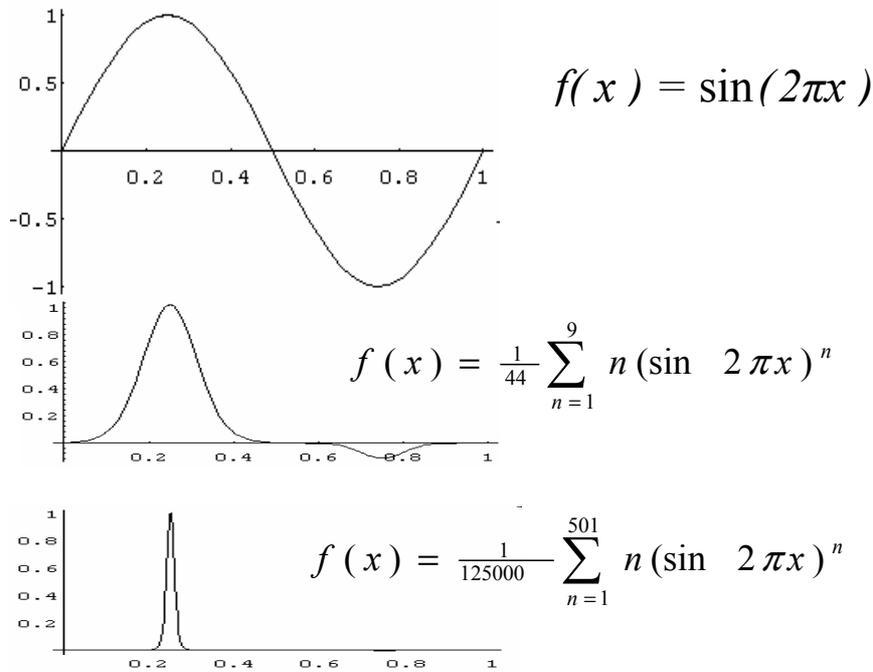

$$f(x) = \sin(2\pi x)$$

$$f(x) = \frac{1}{44} \sum_{n=1}^{9} n (\sin 2\pi x)^n$$

$$f(x) = \frac{1}{125000} \sum_{n=1}^{501} n (\sin 2\pi x)^n$$

**Fig. 1: Higher powers of sinus enable strong localizations**

Quantum theory enables to construct localized phenomena from very extended starting entities. This is a central aspect in this theory, and the postulate is no longer valid that the spacelike small things are also the simplest. On the contrary, the simplest thing that can be imagined, a qubit, is maximally extended over the cosmic space.

Also the difference of a qubit to a quantum field is visible in a clear way.

For a quantum field it is possible to define freely the value of the field strength at any point. The functions in the representation space of a qubit are fixed by the value in one single point at its maximum. This difference explains that with quantum bits the large error, which appears in quantum field theoretical calculations of the cosmological term, does not occur.

For a cosmos with N qubits the first question is: which wavelengths could be expected by physical reasons? The answer follows from the reduction of the tensor product of N two-dimensional representations $D_{1/2}$ of SU(2).[24] For the Clebsch-Gordan series follows: (define $|N/2| = k$ for $N = 2k$ or $N = 2k+1$).

$$D_{\frac{1}{2}}^{\otimes N} = \bigoplus_{j=0}^{|N/2|} \frac{N!(N+1-2j)}{(N+1-j)!j!} D_{|N/2|-j}$$

The factor of the multiplicities

$$f(j) = \frac{N!(N+1-2j)}{(N+1-j)!j!}$$

points out how many functions with which wavelengths can be expected. The multiplicity of the largest wavelength, belonging to $D_0$ and $D_{1/2}$ respectively and therefore to $j=|N/2|$, are of the order of magnitude

$$f(N/2) = \mathcal{O}(2^N N^{-3/2})$$

For shorter wavelengths the multiplicities are growing. The maximum is reached at

$$j \approx \tfrac{1}{2}(N - \sqrt{N})$$

with the order of magnitude

$$f\left(\tfrac{1}{2}(N-\sqrt{N})\right) = O\left(2^N N^{-1}\right) = \sqrt{N} \cdot O\left(2^N N^{-3/2}\right)$$

After this nearly linear growing, an exponential decrease follows. This means that states with an essentially smaller wavelength do not appear in the physical experience.

If $R$ is the curvature radius of the $\mathcal{S}^3$, then the smallest length which can be physically realized with $N$ qubits, is of the order of magnitude

$$\lambda_0 = \frac{R}{\sqrt{N}}$$

---

[24] Görnitz, 1988

The state function, which represents a qubit, therefore has a wavelength of the order

$$R = \sqrt{N} \cdot \lambda_o .$$

If the "energy" of an quantum object is inverse to its extension, then the energy of a qubit is of the order

$$1/\sqrt{N}$$

The total energy $U$ of the $N$ qubits then amounts to

$$N \cdot (1/\sqrt{N}) = \sqrt{N}$$

The volume of the space $\mathcal{S}^3$ is $2\pi^2 R^3$. The energy density $\mu$ in the whole space is proportional to $\sqrt{N} / 2\pi^2 R^3$. We normalize $U$ in such a way that $U = 2\pi^2 R$ and therefore $\mu = 1/R^2$

From the proposed validity of the first law of thermodynamics

$$dU + p \, dV = 0$$

it follows

$$dR + p \, 3 R^2 \, dR = 0$$

or

$$p = -1/3R^2 = -\mu/3$$

From the negative pressure it follows that such a system cannot be static.[25] On the other hand, this pressure fulfills all the energy conditions that are required by physical reasons.[26]

The premise of a distinguished velocity is used for the expansion of the $\mathcal{S}^3$ and therefore also for a definition of the unit for the cosmic time $t$:

$$R = c \, t$$

From our quantum theoretical and thermodynamical considerations in connection with the four premises a cosmological model is derived without any reference to gravitation.

The resulting cosmos is a homogeneous and isotropic $\mathcal{S}^3$, expanding with velocity of light. Its energy-momentum-tensor is

$$T_{ik} = const \cdot diag \, (\mu, p, p, p) = (1/R^2, -1/3R^2, -1/3R^2, -1/3R^2) \qquad (*)$$

---

[25] Landau, Lifschitz, 1971, S. 47
[26] Hawking, Ellis, 1973

and its Friedman-Robertson-Walker-metric can be written as

$$ds^2 = (ct)^2 [(1 - r^2)^{-1} dr^2 + r^2 d\Omega^2] - dt^2$$

The cosmic pressure and energy density fulfill the weak energy condition[27]

$$\mu \geq 0 \text{ und } \mu + p \geq 0$$

with the result, that any observer in its rest system will always measure a positive energy density. Also the dominant energy condition is fulfilled:

$$\mu \geq 0 \text{ und } \mu \geq p \geq -\mu$$

Therefore the velocity of an energy flow, respectively the velocity of sound, is never greater then the velocity of light. Eventually, the strong energy condition

$$\mu + 3p \geq 0 \text{ and } \mu + p \geq 0$$

has the consequence that gravitation is always attractive.

**A solution to the cosmological problems**

Beside the problem of empirical foundation, some other cosmological problems are well known. To these belong the horizon problem, the flatness problem, the problem of the cosmological constant and the "dark energy".

The horizon problem refers to the fact that the cosmic background radiation is almost identical in any direction, whereas in most of the cosmological models no causal contact was possible between the different directions. For its solution the so-called inflation was proposed, but its state equation $\rho = -p$ violates conditions which are necessary for any substance by physical reasons.[28]

The Protyposis-based cosmological model is free of these problems. Because this cosmos expands with the velocity of light, no area exists which before was causally unconnected with other areas.

The problem of the "cosmological constant" is the missing reason for the existence of a quantity which has the value $10^{-122}$ in Planck's units, but which is nevertheless different from zero. Einstein put it into his equation to force the solution to become an everlasting cosmos. Later around the 1980ies, it was decided to set $\Lambda$ equal to zero, but in the last years it became apparent that this leads to serious contradictions to the observational results.

---

[27] Hawking, Ellis, 1973
[28] Hawking, Ellis, 1973

A qubit of the Protyposis is an extremely nonlocal structure. In contradiction to it an object in physics is such a structure that can be localized in space and time and therefore also moved. Under a motion only the state should be changed, but not the object itself. This conception can be defined in Minkowski-space in a clear mathematical way. So, for Minkowski-space a strong definition for a particle exists. A particle in physics is a structure whose states are described by an irreducible representation of the Poincaré-group. Therefore the elementary objects are characterized by its spin, and they can be massless or have a restmass. The vacuum state in Minkowski space and all such particles can be constructed from the qubits of the Protyposis.

The cosmological models can have as essential input dust (i.e. pressureless matter), massless radiation, $\Lambda$ (the cosmological "constant"), and, in the last years, hypothetical dark matter and dark energy.

Parts of the Protyposis can shape into matter and radiation. If the energy-momentum-tensor of the Protyposis is divided into matter and radiation, then it is necessary to introduce a further term with the properties of $\Lambda$. But this tensor is not forced to be constant in time and will not be constant. A further part of the Protyposis is possible, denoted by $\delta$, which can also be adjudicated to the properties of the dark energy. For the dark energy there are no needs for an appearance as "particles".

$$T_{ik} = {}_{(dust)}T_{ik} + {}_{(light)}T_{ik} + {}_{(vacuum)}T_{ik} + {}_{(dark\ energy)}T_{ik}$$

$$\begin{bmatrix} \mu & & & \\ & \frac{\mu}{3} & & \\ & & \frac{\mu}{3} & \\ & & & \frac{\mu}{3} \end{bmatrix} = \begin{bmatrix} \mu_{matter} & & & \\ & 0 & & \\ & & 0 & \\ & & & 0 \end{bmatrix} + \begin{bmatrix} \mu_{light} & & & \\ & \frac{-\mu_{light}}{3} & & \\ & & \frac{-\mu_{light}}{3} & \\ & & & \frac{-\mu_{light}}{3} \end{bmatrix} + \begin{bmatrix} \lambda & & & \\ & \lambda & & \\ & & \lambda & \\ & & & \lambda \end{bmatrix} + \begin{bmatrix} \delta \cdot \mu & & & \\ & \delta \cdot \frac{\mu}{3} & & \\ & & \delta \cdot \frac{\mu}{3} & \\ & & & \delta \cdot \frac{\mu}{3} \end{bmatrix}$$

If $\omega$ is the ratio between the energy density of matter and radiation to the energy density of the vacuum: $\omega = (\mu_{(matter)} + \mu_{(light)}) / \lambda$

Then it follows[29]

$$\lambda = (1-\delta)\ \mu / (\omega + 1)$$

$$\mu_{(light)} = (1-\delta)\ \mu\ (2 - \omega) / (\omega + 1)$$

---

[29] Görnitz, 1988

$$\mu_{(matter)} = (1-\delta) \, 2\mu(\omega-1)/(\omega+1).$$

Since $\mu > 0$, $\mu_{(matter)} > 0$, $\mu_{(light)} > 0$, it follows $2 \geq \omega \geq 1$

This ansatz solves the problem of the cosmological term. The magnitude of $\lambda$, which has the properties of the energy density of the vacuum, has to be of the same order as the energy density of matter and radiation, and not, like in quantum field theoretical computations, many orders of magnitude larger. The flatness problem is solved by the large cosmic curvature radius of the $\mathcal{S}^3$ for a cosmos expanding with velocity of light since 14 billions of years.

**Comparison with observations**

A comparison with the observations are at present more encouraging than in the 1980ies. For the relation between Hubble-parameter $H$ and cosmic time $t$, the results of the supernovae-data show for the present values (index 0): $H_0 \, t_0 = 0.96 \pm 0.04.$

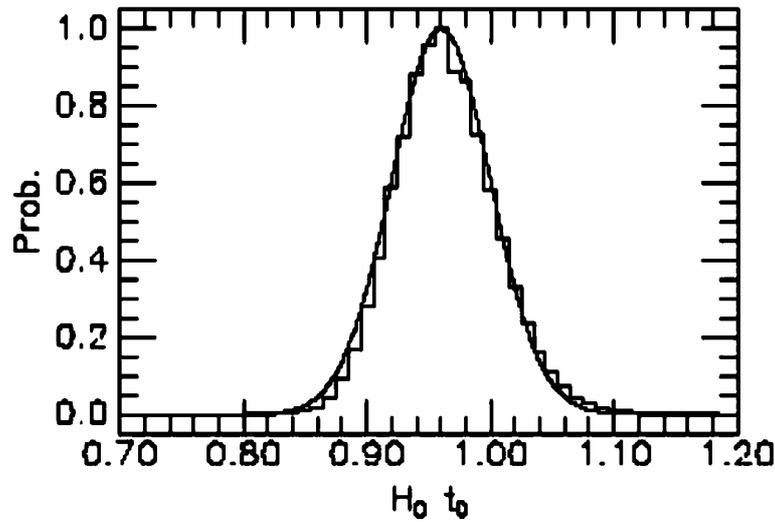

**Fig. 2: Relation between Hubble-parameter and cosmic time**[30]

The same data, $t_0 = 13.75 \pm 0.13$ Gyr $= 433{,}62 \cdot 10^{15}$ s and $H_0 = 71.0 \pm 2.5$ km/s/Mpc, so $1/H_0 = 0{,}4336 \cdot 10^{18}$ s, can be found by Larson et al.[31] The Protyposis-model claims that the value *1* is not an accident of the present time, moreover, it should hold in general

$$H \, t = 1.$$

---

[30] Tonry et al, (2003), Fig. 15
[31] Larson 2010

At present a picture of the cosmos is preferred in which the cosmic space expands with an accelerating velocity. But the data from the supernovae observations seem to be reconcilable with the Protyposis-model as well.

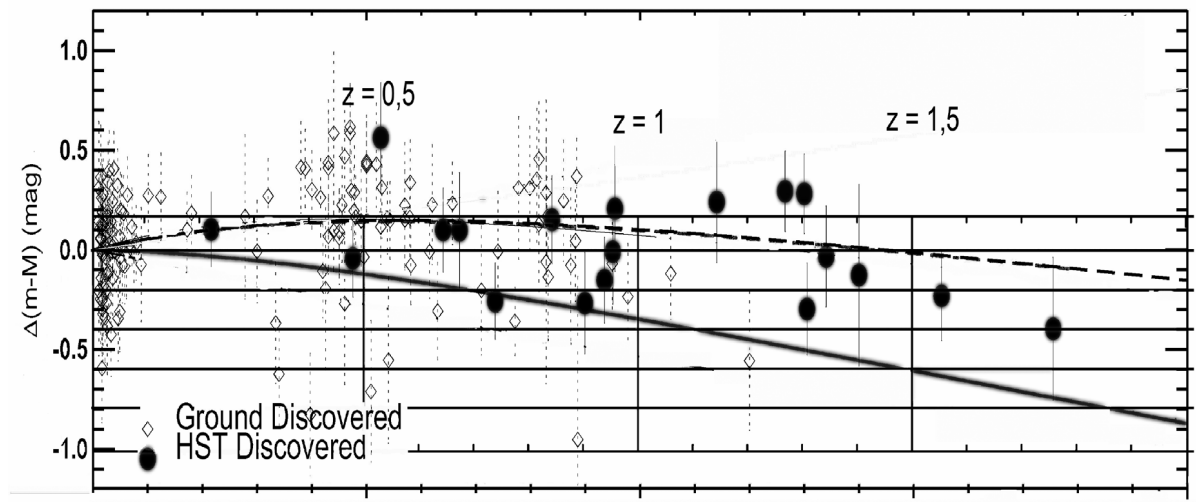

**Fig. 3: Comparison of the supernovae data[32] with the model of an accelerated expanding cosmos (- - -) and with a cosmos expanding with velocity of light (Protyposis-model) (——)[33]**

From the presented considerations it is to expect that the tendency of the data will increase, that for larger values of z, i.e. for large scales, the data approach the lower one of the both curves. The "Binned Gold data" from Riess et.al.[34] seem to show such an effect. My expectation is underpinned by an argument of Aurich, Lustig und Steiner[35]: „but there remains a strange discrepancy at large scales as first observed by COBE[36] and later substantiated by WMAP[37]." and further: „The suppression of the CMB anisotropy at large scales respectively low multipoles can be explained if the universe is finite."

For a clarification of the connection of the Protyposis-cosmology with the structure of the background radiation further investigations are necessary because Aurich, Lustig und Steiner also denote that with the assumptions that they have proposed, a simply connected $\mathcal{S}^3$ does not fit the data very well, therefore, they have used more complicated topologies.

---

[32] Riess et. al. , 2004, Fig. 7
[33] Görnitz, 2006, S. 163 ff
[34] Riess et. al. 2007, Fig. 6
[35] Aurich, Lustig, Steiner, 2005
[36] Hinshaw et al., 1996
[37] Bennett et al., 2003

## A foundation of general relativity from the Protyposis

The Protyposis model leads to a cosmological model with a distinguished background metric. Already Dirac has referred to the fact that such a distinguished background metric remains unnoticeable as long as one is included into an "Einsteinian elevator" without any windows. The situation changes if a window to the cosmos is opened and the background radiation is included into the researches.[38] That all reference systems are of equal rights is therefore a local property in the description of localized systems with arbitrary accelerations, however, it is not necessary to claim this property for the whole cosmic space as well.

If the metric of the Protyposis model

$$ds^2 = -dt^2 + (ct)^2 [(1 - r^2)^{-1} dr^2 + r^2 d\Omega^2]$$

is compared with the energy-momentum-tensor $T_{ik}$ in (*), no relation is evident at a first sight.

But it is possible to put, as usual, $c = 1$, and to re-write it

$$ds^2 = -dt^2 + t^2 [d\chi^2 + \sin^2\chi(d\theta^2 + \sin^2\theta\, d\varphi^2)] ,$$

to introduce the orthonormal basis of one-forms[39]

$$\omega^t = dt \quad \omega^\chi = t\, d\chi \quad \omega^\theta = t \sin\chi\, d\theta \quad \omega^\varphi = t \sin\chi \sin\theta\, d\varphi ,$$

and to compute the Einstein-tensor $G_{ik}$ (the curvature-constant for this closed space is $k=+1$):

$G_{tt} = 3(a_{,t}/a)^2 + 3k/a^2 = 3(1/t)^2 + 3/t^2 = 6/t^2$

$G_{\chi\chi} = G_{\theta\theta} = G_{\varphi\varphi} = -2\, a_{,tt}/a - (a_{,t}/a)^2 - k/a^2 = 0 - (1/t)^2 - 1/t^2 = -2/t^2$

In this basis the energy-momentum-tensor $T_{ik}$ is

$T_{tt} = \rho = const/R^2,$  $\qquad T_{\chi\chi} = T_{\theta\theta} = T_{\varphi\varphi} = p = - const/(3R^2),$

and with $R = t$

it follows $\qquad\qquad\qquad G_{ik} = 6\, const\, T_{ik}$

It becomes evident that this $G_{ik}$ is proportional to the tensor $T_{ik}$.

If it is demanded that this proportionality remains valid also for local variations of the energy density and the pressure, which are for large scales in a good approximation homogeneous and isotropic, the general relativity remains valid as a very good description of local inhomogeneities of cosmological spacetime.

---

[38] Dirac, 1980
[39] Misner, Thorne, Wheeler 1973, pp. 728

The cosmological expansion can be understood as an expression for the growing of the amount of quantum information. The rising quantity of quantum information enables an more and more finer division of space and also an evolution of even more complex structures in the cosmic space. Such structures disturb the homogeneity of space, and gravity can be understood in a metaphorical way as the reaction of the cosmos on such variations.

**Concluding remarks**

There is a long line of papers from Bekenstein and Hawking over Jacobson until Verlinde and Hossenfelder, to mention only very few of them, in which the connection between entropy and gravity is examined. The way from Weizsäcker's Ur-Alternatives to the Protyposis opens the view on an even larger field. If in addition to entropy all of the quantum information is taken into account then beside gravitation also a new understanding of matter becomes possible. It reveals, adjacent to the insoluble interconnection of quantum theory and cosmology, also the conjunction of these areas of science to the wide field of the mind. But this is performed elsewhere.[40]

**Literatur**

Aurich, R., Lustig, S., Steiner, F. (2005): CMB Anisotropy of Spherical Spaces, *Class. Quant. Grav*. **22,** 3443-3460

Bekenstein. J. D. (1973): Black holes and entropy, *Physical Review* **D 7**, 2333-2346

Bekenstein, J. D. (1974): Generalized second law of thermodynamics in black hole physics, *Physical Review* **D 9,** 3292-3300

Bennett, C. L. et al. (2003): FIRST-YEAR WILKINSON MICROWAVE ANISOTROPY PROBE (WMAP)1 OBSERVATIONS, *Astrophys. J. Supp*. **148,** 1

Dirac, P.A.M. (1980): "WHY WE BELIEVE IN THE EINSTEIN THEORY", in Gruber, B., Millman, R. S. (Eds.): *SYMMETRIES IN SCIENCE*, Plenum Press, New York, London,

Drieschner, M. (1979): „V*oraussage Wahrscheinlichkeit Objekt"*, Lecture Notes in Physics **99**, Springer, Berlin,

Einstein, A. (1984): „*Aus meinen späten Jahren"*, Frankfurt a. M., Berlin, Wien: Ullstein, p. 90

---

[40] Görnitz&Görnitz 2008

Finkelstein, D.: (1968) Space-time code, *Phys. Rev.* **185**, 1261

Goenner, H. (1994): „*Einführung in die Kosmologie*", Heidelberg, Spektrum, p. 87

Görnitz, Th. (1988a): Abstract Quantum Theory and Space-Time-Structure, Part I: Ur-Theory, Space Time Continuum and Bekenstein-Hawking-Entropy, *Intern. Journ. Theoret. Phys.* **27,** 527-542

Görnitz, Th. (1988b): On Connections between Abstract Quantum Theory and Space-Time-Structure, Part II: A Model of cosmological evolution, *Intern. Journ. Theoret. Phys.* **27,** 659-666

Görnitz, Th., Graudenz, D., Weizsäcker, C. F. v. (1992): Quantum Field Theory of Binary Alternatives; *Intern. J. Theoret. Phys.* **31**, 1929-1959

Görnitz, Th., Schomäcker, U. (1996): „Zur Beziehung zwischen Quantentheorie und klassischer Physik", Talk given at GROUP 21, Goslar,

Görnitz, Th. (1999): „*Quanten sind anders*", Heidelberg, Spektrum,

Görnitz, Th. (2006): *"Cosmology and particle Physics"*, in Görnitz, Th., Lyre, H. (Eds.): „*C.F.v. Weizsäcker – Structure of Physics*", Springer, Berlin, pp. 150-178

Görnitz, Th. & Görnitz, B. (2008): "*Die Evolution des Geistigen*", Göttingen, Vandenhoeck & Ruprecht,

Görnitz, Th. (2009): From Quantum Information to Gravitation, arXiv:0904.1784

Hawking, S. W., Ellis, G. F. R. (1973): "*The Large Scale Structure of the Universe*", University Press, Cambridge

Hawking, S. W. (1975): Particle creation by black holes, *Communications in Mathematical Physics* **43,** 199-220.

Hinshaw, G. et al. (1996): Two-Point Correlations in the *COBE*[*] DMR Four-Year Anisotropy Maps *Astrophys. J. Lett.* **464**, L25

Hossenfelder, S.: Comments on and Comments on Comments on Verlinde's paper "On the Origin of Gravity and the Laws of Newton" arXiv:1003.1015v1

Jacobson, T. (1995): Thermodynamics of Spacetime: The Einstein Equation of State, Phys. Rev. Lett. **75**, 1260-1263

Landau, L. D., Lifschitz, E. M. (1971): „*Lehrbuch der Theoretischen Physik, Bd. V.*", Akademie-Verlag, Berlin, p. 47

Larson, D. et. al. (2010): SEVEN-YEAR WILKINSON MICROWAVE ANISOTROPY PROBE (WMAP1) OBSERVATIONS: POWER SPECTRA AND WMAP-DERIVED

PARAMETERS, arXiv:1001.4635v2

Padmanabhan, T. (2010): Thermodynamical Aspects of Gravity: New insights, arXiv:0911.5004v2

Riess, A. G. et. al. (2004): Type Ia Supernova Discoveries at z > 1 From the Hubble Space Telescope: Evidence for Past Deceleration and Constraints on Dark Energy Evolution, *Astrophys. J.* **607,** 665-687

Riess, A. G. et. al. (2007): NEW HUBBLE SPACE TELESCOPE DISCOVERIES OF TYPE Ia SUPERNOVAE AT z ≥ 1: NARROWING CONSTRAINTS ON THE EARLY BEHAVIOR OF DARK ENERGY, *Astrophys. J.* **659**,98-121

Scheibe, E., Süssmann, G., Weizsäcker C. F. v. (1958): Mehrfache Quantelung, Komplementarität und Logik III, *Zeitschrift für Naturforschung*, **13a**, 705.

Schommers, W. (2008): Cosmological Constant, Space-Time, and Physical Reality, *Advanced Science Letters*, **1**, 59–91

Tonry, J. L. et al. (2003): Cosmological Results from High-z Supernovae, *Astrophysical J.* **594**,1-24

Verlinde, E. (2010): On the Origin of Gravity and the Laws of Newton, arXiv:1001.0785v1

Wheeler, J. A. (1990): "Information, physics, quantum: The search for links", in: W. Zurek, ed., *Complexity, Entropy, and the Physics of Information*, Redwood City, CA: Addison-Wesley

Weizsäcker, C. F. v. (1955): Komplementarität und Logik I, *Naturwissenschaften*, **42**, 521-529, 545-555

Weizsäcker, C. F v. (1958): Komplementarität und Logik II, *Zeitschrift für Naturforschung*, **13a,** 245

Weizsäcker, C. F. v. (1971): „*Die Einheit der Natur"*, Hanser, München, [Engl. Ed.: *The Unity of Nature*, Farrar. Straus, Giroux, NewYork, 1980].

Weizsäcker, C. F. v. (1985): "*Aufbau der Physik*", Hanser, München, [Engl. Ed.: Görnitz, Th., Lyre, H. (Eds.): „*C. F. v. Weizsäcker – Structure of Physics*", Springer, Berlin, 2006]

,